\documentclass[sigconf]{aamas}

\usepackage{balance} 
\usepackage{graphicx}
\usepackage{adjustbox}
\usepackage{amsmath}
\usepackage{bm}
\usepackage{bbm}
\usepackage{mathtools}

\usepackage{amsthm}
\usepackage[acronyms, shortcuts]{glossaries}

\usepackage{wrapfig}
\usepackage{subcaption}
\usepackage{lipsum}

\usepackage{siunitx}
\usepackage{xcolor}
\usepackage[ruled,vlined,linesnumbered]{algorithm2e}
\usepackage{ifthen}
\usepackage{dsfont}
\usepackage{dirtytalk}
\usepackage{afterpage}

\usepackage{times}
\usepackage{multirow}

\usepackage[T1]{fontenc}

\usepackage[capitalize,noabbrev,nameinlink]{cleveref}
\crefformat{equation}{(#2#1#3)}
\Crefname{algocfline}{Algorithm}{Algorithms}
\Crefname{algocf}{line}{lines}

\newtheorem{assumption}{Assumption}
\newtheorem{proposition}{Proposition}

\allowdisplaybreaks

\newcommand{\cost}{J}
\newcommand{\conseq}{G}
\newcommand{\consineq}{H}

\newcommand{\var}{x}
\newcommand{\vars}{\mathbf{\var}}
\newcommand{\feasibleset}{\mathcal{X}}
\newcommand{\dimvar}{n}
\newcommand{\numplayers}{N}
\newcommand{\dimworlds}{m}
\newcommand{\signal}{\sigma}
\newcommand{\signals}{\mathcal{S}}
\newcommand{\world}{\omega}
\newcommand{\worlds}{\Omega}
\newcommand{\struct}{r}
\newcommand{\structs}{\mathbf{r}}

\newcommand{\structcost}{K}

\newcommand{\preference}{\beta}
\newcommand{\preferences}{\mathbf{B}}
\newcommand{\difference}{\delta}
\newcommand{\activation}{\zeta}
\newcommand{\prior}{p}
\newcommand{\perturbation}{\theta}
\newcommand{\gradkkt}{\mathbf{F}}
\newcommand{\solution}{\mathbf{z}}

\newcommand{\varbrev}{\mathbbmss{\var}}

\newcommand{\fernando}[1]{{\textcolor{purple}{\textsc{[Fernando: {#1}]}}}}

\newacronym{cig}{CIG}{Complete Information Game}
\newacronym{gameincomplete}{DIIG}{Defender Incomplete Information Game \fernando{change this}}
\newacronym{gamesignal}{SPICE}{\underline{s}mooth, two-\underline{p}layer game with asymmetric and \underline{i}n\underline{c}ompl\underline{e}te information}
\newacronym{mcp}{MCP}{mixed-complementarity problem}
\newacronym{cp}{CP}{certain player}
\newacronym{up}{UP}{uncertain player}

\setcopyright{ifaamas}
\acmConference[AAMAS '25]{* denotes equal contribution. Proc.\@ of the 24th International Conference
on Autonomous Agents and Multiagent Systems (AAMAS 2025)}{May 19 -- 23, 2025}
{Detroit, Michigan, USA}{A.~El~Fallah~Seghrouchni, Y.~Vorobeychik, S.~Das, A.~Nowe (eds.)}
\copyrightyear{2025}
\acmYear{2025}
\acmDOI{}
\acmPrice{}
\acmISBN{}

\acmSubmissionID{1029}

\title[Smooth Information Gathering in Two-Player Noncooperative Games]{Smooth Information Gathering in Two-Player Noncooperative Games}

\author{Fernando Palafox}
\affiliation{
  \institution{University of Texas at Austin}
  \city{Austin, TX}
  \country{United States}}
\email{fernandopalafox@utexas.edu}

\author{Jesse Milzman}
\affiliation{
  \institution{DEVCOM Army Research Laboratory}
  \city{Adelphi, MD}
  \country{United States}}
\email{jesse.m.milzman.civ@army.mil}

\author{Dong Ho Lee}
\affiliation{
  \institution{University of Texas at Austin}
  \city{Austin, TX}
  \country{United States}}

\author{Ryan Park}
\affiliation{
  \institution{University of Texas at Austin}
  \city{Austin, TX}
  \country{United States}}

\author{David Fridovich-Keil}
\affiliation{
  \institution{University of Texas at Austin}
  \city{Austin, TX}
  \country{United States}}

\begin{abstract}
We present a mathematical framework for modeling two-player noncooperative games in which one player is uncertain of the other player's costs but can preemptively allocate information-gathering resources to reduce this uncertainty.
We refer to the players as the \ac{up} and the \ac{cp}, respectively.
We obtain \ac{up}'s decisions by solving a two-stage problem where, in Stage 1, \ac{up} allocates information-gathering resources that smoothly transform the information structure in the second stage.
Then, in Stage 2, a signal (that is, a function of the Stage 1 allocation) informs \ac{up} about \ac{cp}'s costs, and both players execute strategies which depend upon the signal's value.
This framework allows for a smooth resource allocation, in contrast to existing literature on the topic. 
We also identify conditions under which the gradient of \ac{up}'s overall cost
with respect to the information-gathering resources is well-defined.
Then we provide a gradient-based algorithm to solve the two-stage game. 
Finally, we apply our framework to a tower-defense game which can be interpreted as a variant of a Colonel Blotto game with smooth payoff functions and uncertainty over battlefield valuations.
We include an analysis of how optimal decisions shift with changes in information-gathering allocations and perturbations in the cost functions. 
\end{abstract}


\newcommand{\BibTeX}{\rm B\kern-.05em{\sc i\kern-.025em b}\kern-.08em\TeX}

\begin{document}


\pagestyle{fancy}
\fancyhead{}


\maketitle 


\section{Introduction}

\begin{figure*}
    \centering
    \includegraphics[width=0.80\textwidth]{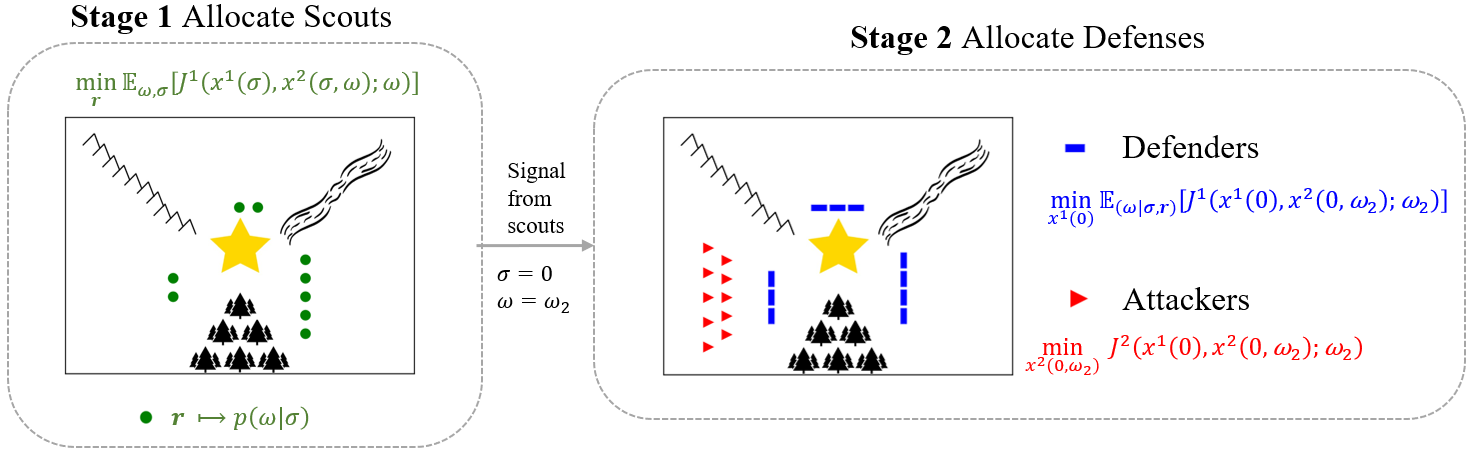}
    \caption{Overall schematic of the two-stage game applied to a tower defense scenario. 
    In this scenario, \ac{up} must defend a tower (denoted by a star) that can be attacked from three directions. 
    In Stage 1, \ac{up} seeks to minimize its expected cost by allocating information-gathering resources $\structs$, and thereby defining posterior distribution $p(\world|\signal)$. Here, ``world'' $\world \in \{\world_1, \world_2, \world_3\}$ reflects the direction which \ac{cp} wishes to attack, and ``signal'' $\signal \in \{0, 1, 2, 3 \}$ is the output of a noisy sensor. In Stage 2, both players seek to minimize their costs $\cost^1,\cost^2$;
    Their allocations $(\var^1, \var^2)$ are functions of the signal $\signal$ (\ac{up}, player 1) and/or the world $\world$ (\ac{cp}, player 2). 
    }
    \label{fig:intro_fig}
\end{figure*}

Incomplete information games provide a mathematical formalism for understanding
the behavior of rational agents who lack perfect knowledge of one another's objectives or other aspects of the game
\cite{Harsanyi1968bayesian}.
In real-world scenarios, agents often attempt to preemptively gather information before such interactions in order to reduce their uncertainty and gain a strategic advantage.
However, any player gaining new information may have to contend with other players shifting their strategies in response.
Traditionally, such rational information-gathering activities 
are 
framed as a discrete choice, e.g., whether or not to pay for costly information with a pre-defined structure \cite{xu2016modeling}.
In contrast, real-world information-gathering decisions, e.g., distributed sensor placement \cite{kaplan2006global}, are often continuously parametrized.

In this paper, we address this question of continuously parameterized, preemptive information-gathering.
To this end, we develop a model for two-player non-cooperative games of incomplete information with the following features:
\begin{itemize}
    \item The \acrfull{up} does not know the \acrfull{cp}'s cost.
    \item \ac{up} can smoothly allocate information-gathering resources to reduce uncertainty, and \ac{cp} is aware of this allocation. 
\end{itemize}
In such games, there is a coupling between the optimal allocation of information-gathering resources and both players' strategic reactions. 

Our contributions are as follows: (1) A game-theoretic model for two-player noncooperative games with one-sided uncertainty and two stages, as shown in \cref{fig:intro_fig}. 
In Stage 1, \ac{up} selects how to allocate information-gathering resources from a continuous decision landscape. 
This allocation parameterizes the relationship between a world unknown to \ac{up}, and a signal that provides \ac{up} with limited information about the world. 
In Stage 2, \ac{up} receives the value of the signal which both players then use to play a non-cooperative game.
(2) A local descent algorithm to solve both Stage 1 and Stage 2 for each player's decisions.
(3) Conditions under which gradients of costs and solutions with respect to decision variables are well-defined.
(4) An application of this model to a tower defense scenario---akin to a Colonel Blotto game with smooth payoff functions---and an analysis of the solutions.

\section{Related Work}


Interactions in which some players are uncertain about others' objectives were first formally modeled in Harsanyi's seminal work on Bayesian games \cite{Harsanyi1968bayesian}. 
Since then, Bayesian games have been applied to a variety of domains including, but not limited to, cybersecurity of nuclear plants \cite{Maccarone2022nuclear}, intrusion detection in wireless networks \cite{liu2006intrusion}, and decision-making in military command and control settings \cite{Brynielsson_Arnborg_2004, hunt2023review}. 
Much of the existing literature focuses primarily on optimal decision-making in these uncertain environments, typically without examining the capacity of agents to transform the information landscape to their advantage. 
A notable exception is the literature on deception, e.g., \cite{xu2016modeling,hespanha2000deception, rostobaya2023deception}.

In single-agent settings, it is common to study this concept of altering the information landscape via the value of information (VoI), which quantifies the expected reduction in cost
attributable to a given source of information---i.e., a given variable available to the decision-maker.
VoI can be understood as the amount a decision-maker would be willing to pay for information before making a decision \cite{Howard1966voi}.
VoI was traditionally related to the comparison of experiments \cite{blackwell1953} from statistical decision theory (SDT) \cite{raiffa1968applied}, which has game-theoretic roots \cite{blackwell1949comparison}.
In recent decades, VoI in Bayesian games has been studied as the comparison of information structures \cite{lehrer2006voi,lehrer2010signaling,pkeski2008comparison,hogeboom2021comparison}, which essentially adapts the SDT framework for experiments to the more complex information-cost interactions that emerge from multiple decision-makers.

In this work, we focus on optimizing preemptive actions that selectively gather information to strategically minimize uncertainty in a noncooperative, Bayesian, two-player game.
Incorporating VoI-based information gathering into a game where one player has the opportunity to rationally manipulate their information structure \textit{ex ante} is a less-studied phenomenon.
To the best of our knowledge, it has not been studied for smooth manipulations of the information structure, as we do in this work.

We highlight a few similar efforts.
In earlier work, \citet{hespanha2000deception} introduced a partial information attacker/defender game in which the defender acts first and may selectively reveal some information about their resource allocation, in order to deceive their opponent.
Fuchs and Khargonekar \cite{fuchs2012} similarly proposed a two-stage, Stackelberg-like partial information Colonel Blotto game.
In this game, one player allocates their resources, and then the second player receives a signal (with a given structure) from a sensor system which alerts them to whether the resources allocated to each battlefield are above a fixed threshold.
The second player then uses this information for their allocation.
However, they do not consider sensor placement (as a resource allocation problem) or individual sensor tuning, i.e., VoI-based decisions on what information structures would be optimal.
More directly comparable to the problem we investigate, Xu and Zhuang \cite{xu2016modeling} set up an incomplete information attacker-defender game, in which the attacker has a costly choice of whether or not to try to learn the vulnerability of the defender---a binary variable chosen by nature.
This leads to a characterization of the conditions under which the attacker benefits from trying to learn the defender's vulnerability.

To the best of our knowledge, no existing literature has explicitly modeled the pre-emptive acquisition of information as a continuously-parametrized decision on the part of one of the players, prior to strategic engagement.
Insofar as many real-world informational decisions are smoothly parametrized---e.g. the placement, orientation, and calibration of stationary sensors, or the utilization of drones to conduct reconnaissance along spatially and temporally continuous paths---there is a need for the development of theory and methods for optimizing smooth information structures for decision-making in non-cooperative settings, adversarial or otherwise.
This work is an initial effort in that direction.
\section{Formulation}

We begin by defining a noncooperative complete information game and build on it to formulate a Bayesian game with asymmetric uncertainty about \ac{cp}'s objective. 
Then, we introduce the concept of signals and signal structure in order to model \ac{up}'s information-gathering measures and define a smooth, two-player game with asymmetric and incomplete information.  
Finally, we formulate the problem of optimally allocating information-gathering as a two-stage problem, in which the continuous decision of the first stage smoothly transforms the information structure in the second stage.

Throughout the paper, superscripts denote indexing by a player, where $1$ and $2$ denote the uncertain (\ac{up}) and certain (\ac{cp}) players, respectively. 
Subscripts denote indexing elements in a vector. 
We use the following notation to refer to sets $[\dimvar] \equiv \{1, \hdots, \dimvar\}$.

\subsection{Complete Information Game}

We begin by defining two-player, static games with complete information.
Each player $i$ seeks to minimize a cost function $\cost^i: \mathbb{R}^{2  \dimvar} \to \mathbb{R}$ subject to constraints $\conseq^i(\var^i) = 0$ and $\consineq^i(\var^i) \geq 0$, where $\conseq^i: \mathbb{R}^{\dimvar} \to \mathbb{R}$, $\consineq^i: \mathbb{R}^{\dimvar} \to \mathbb{R}$, 
and $\dimvar$ is the dimension of each player's action space.

Mathematically, each player's problem is given by 
\begin{subequations}
\label{adversarial_game}
\begin{align}
&\min_{\var^1} \cost^{1}(\var^1, \var^2) \mathrm{~s.t.~} \var^1 \in \feasibleset^1\\
&\min_{\var^2} \cost^{2}(\var^1, \var^2) \mathrm{~s.t.~} \var^2 \in \feasibleset^2
\end{align}
\end{subequations}
where $\feasibleset^i = \{\var^i ~|~ \conseq^i(\var^i) = 0, \consineq^i(\var^i) \geq 0 \}$ is player $i$'s feasible set. 
Note that the players' decisions $\var^i$ are coupled via $\cost^i(\cdot).$

\subsection{Bayesian Game With Asymmetric Uncertainty}
\label{sec:asymmetric}
Now we extend the game to include \ac{up}'s uncertainty about \ac{cp}'s cost, 
where \ac{up} must make a decision without any new information but its prior knowledge.
We introduce uncertainty by parameterizing the costs with an unknown ``world'' denoted $\world \in \worlds = [\world_\dimworlds]$, where $\dimworlds$ is the number of worlds. 
We assume \ac{up} has a prior belief about what the world is in the form of a discrete distribution $p: \worlds \to [0,1], \; \sum_{\omega} p(\omega) = 1$. 
Since \ac{up} is uncertain about the value of $\world$, it now seeks to minimize its expected cost over the prior distribution $p(\world)$. 
However, computing this expectation requires knowledge of \ac{cp}'s decisions for every possible world, i.e., $\var^2(\world)$ for every $\world \in \worlds$. 
Therefore, the new game is given by 
\begin{subequations}
\label{game_incomplete}
\begin{alignat}{3}
    &~\min_{\var^1} ~\mathbb{E}_\world[\cost^{1}(\var^1, \var^2(\world); \world)]\quad  &&\mathrm{~s.t.~} \var^1 \in \feasibleset^1\\
    &\min_{\var^2(\world_1)} \cost^{2}(\var^1, \var^2(\world_1); \world_1) &&\mathrm{~s.t.~} \var^2(\world_1)\in \feasibleset^2\\
    &\qquad \qquad \vdots \notag\\
    &\min_{\var^2(\world_\dimworlds)} \cost^{2}(\var^1, \var^2(\world_\dimworlds); \world_\dimworlds) &&\mathrm{~s.t.~} \var^2(\world_\dimworlds)\in \feasibleset^2.
\end{alignat}
\end{subequations}

Note that the information structure of \cref{game_incomplete} implies that \ac{cp} is aware of \ac{up}'s uncertainty since all of \ac{cp}'s types $\var^2(\world), \world \in \worlds$ play against a single \ac{up} $\var^1$, a decision made on the basis of the common prior $p(\omega)$.

\subsection{Signals and Signal Structures}
\label{subsection:signals}

Our next step is to provide a mathematical framework that describes \ac{up}'s capacity to deploy information-gathering measures, e.g., by deploying surveillance resources.
To that end, we introduce the concept of a signal and a signal structure. 
Before deciding on the value of $\var^1$, suppose that \ac{up} receives a signal $\signal \in \signals = \{0, 1, \dots, \dimworlds\}$ with information about the true value of $\world$.
We associate one signal value $k$ for each world $\omega_k$, with a signal of $0$ signifying an information-gathering failure. 
The relationship between signal $\signal_i$ and world $\world_j$ is determined by the signal structure: the conditional probability $p(\signal_i | \world_j)$, which will be determined by \ac{up}'s information-gathering decision. 

We make the following assumptions regarding the signal structure:
\begin{assumption}[No false positives]
\label{assumption:no_false_positives}
$p(i | \world_j) = 0 ~\forall i \neq j, i>0$. This implies $p(0|\world_i) = 1 - p(i|\world_i)~ \forall \world_i \in \worlds$.
\end{assumption}
\begin{assumption}[\ac{cp}'s awareness]
 \label{assumption:awareness}
\ac{cp} is aware of both the signal value and signal structure. 
\end{assumption}

From \cref{assumption:no_false_positives} it follows that signal $i>0$ always implies that $\world = \world_i$. 
\cref{assumption:awareness} adds information asymmetry to the interaction: not only is \ac{up} uncertain about \ac{cp}'s true intentions, but \ac{cp} is also aware of the signal structure (e.g., surveillance allocation) and the received signal value.
This models a worst-case scenario where \ac{cp} is fully aware of the information available to the \ac{up}. 

\subsection{Smooth, Two-Player Game with
Asymmetric and Incomplete Information}
\label{sec:smooth}

We now introduce a two-player game that encodes the knowledge gained from the signal structure defined in the previous section.
\ac{up}'s prior distribution remains unchanged. 
In this game, \ac{up} makes their decision using the signal value, i.e., $\var^1(\signal)$, and \ac{cp} makes a decision using both the signal and the world value, i.e., $\var^2(\signal, \world)$.
Thus, we are solving the Bayesian game given by
\begin{subequations}
\label{game_signal_compact}
\begin{align}
    &\min_{\var^1} \mathbb{E}_{\world, \signal}[\cost^{1}(\var^1(\signal), \var^2(\signal,\world);\world)]\label{eq:game_signal_compact_a}\\
    &\min_{\var^2} \mathbb{E}_{\world, \signal} [\cost^{2}(\var^1(\signal), \var^2(\signal,\world); \world)]\\
    \intertext{where the player strategies are maps of the form}
    &\qquad \qquad  \var^1: \signals \to \feasibleset^1 \\
    &\qquad \qquad  \var^2: \signals \times \worlds \to \feasibleset^2.
\end{align}
\end{subequations}

\cref{assumption:no_false_positives} implies that \ac{up}'s decision when $\signal \neq 0$ does not depend on its prior $p(\omega)$, since both players know $\omega$. 
However, if $\signal = 0$, $\world$ could still be any $\world_i$ for which $p(0|\world_i) > 0$. 
Therefore, to select $\var^1(0)$, \ac{up} must minimize the expectation of its cost over the conditional probability $p( \world_i|0)$ for every $\world_i$ such that $p(0|\world_i) > 0$.
Using this information, we may break up the game given by (\ref{game_signal_compact}) into its component decisions, given by:
\begin{subequations}
\label{game_signal}
\begin{alignat}{2}
    &~\min_{\var^1(0)} &&\mathbb{E}_{\world|0}[\cost^{1}(\var^1(0), \var^2(0,\world);\world)]\label{eq:game_signal_a}\\
    &~\min_{\var^1(1)} &&\cost^{1}(\var^1(1), \var^2(1,\world_1);\world_1)\label{eq:perfect_info_p1_start}\\
    & &&\quad \vdots \notag\\
    &~\min_{\var^1(\dimworlds)} &&\cost^{1}(\var^1(\dimworlds), \var^2(\dimworlds,\world_\dimworlds); \world_\dimworlds)\label{eq:perfect_info_p1_end}\\
    &\min_{\var^2(0, \world_1)} &&\cost^{2}(\var^1(0), \var^2(0,\world_1); \world_1)\label{eq:imperfect_info_start}\\
    & &&\quad\vdots \notag\\
    &\min_{\var^2(0, \world_\dimworlds)} &&\cost^{2}(\var^1(0), \var^2(0,\world_\dimworlds); \world_\dimworlds)\label{eq:imperfect_info_end}\\
    &\min_{\var^2(1, \world_1)} &&\cost^{2}(\var^1(1), \var^2(\dimworlds,\world_1); \world_1)\label{eq:perfect_info_p2_start}\\
    & &&\quad \vdots \notag\\
    &\min_{\var^2(\dimworlds, \world_\dimworlds)} &&\cost^{2}(\var^1(\dimworlds), \var^2(\dimworlds,\world_\dimworlds); \world_\dimworlds),\label{eq:perfect_info_p2_end}
\end{alignat}
\end{subequations}
subject to $\var^1(\signal) \in \feasibleset^1, \forall \signal \in \signals$ and $\var^2(\signal, \world) \in \feasibleset^2, \forall \signal \in \signals, \world \in \worlds$.
Note that \cref{eq:game_signal_a} depends on the signal structure $p(0 | \world_i)$ via Bayes' rule. 
The term $\var^1(0)$ in \cref{eq:game_signal_a} can be interpreted as \ac{up}'s decision given no warning and accounting for its knowledge of how it allocated information-gathering resources. 
For example, consider a situation where \ac{up} allocated enough information-gathering resources to ensure that it will always be warned when $\world = \world_1$, i.e., $p(1|\world_1) = 1$. 
Then, it need not account for $\var^2(0, \world_1)$ when minimizing the expected cost in \cref{eq:game_signal_a} because it knows that $p(0|\world_1) = 0$.

By contrast, the solutions to the complete information subgames given by \cref{eq:perfect_info_p1_start}-\cref{eq:perfect_info_p1_end} and 
\cref{eq:perfect_info_p2_start}-\cref{eq:perfect_info_p2_end} are completely independent of the signal structure.

We remark that this game is a generalization of the Bayesian game described in \cref{sec:asymmetric}. 
To see why, note that setting $p(0 | \world) = 1,~\forall \world \in \worlds$ exactly reduces \cref{game_signal} into \cref{game_incomplete} since the only relevant decisions are those for which $\signal = 0$ (as $p(\signal|\world) = 0,~\forall \signal > 0$).
This corresponds to the case of a \ac{up} without the capability to gather new information. 

\subsection{Signal Structure Selection}

\ac{up} seeks a signal structure that will strategically minimize their expected cost \cref{eq:game_signal_compact_a}. 
To that end, we parametrize the signal structure from Sec.~\ref{subsection:signals} with the decision variable $\structs \in \mathbb{R}_{\geq 0}^\dimworlds, \sum_i \struct_i=1$
such that
$\struct_i = p(\signal_i | \world_i)$.
We may then formulate the signal structure selection problem as 
\begin{subequations}
\label{structure_selection}
    \begin{align}
        &\min_\structs \structcost\label{structure_objective}\\
        &\mathrm{s.t.~} 0 \leq \struct_i \leq 1\label{structure_probability}\\
        &\qquad \sum_{i = 1}^\dimworlds \struct_i = 1\label{structure_limits}
    \end{align}
\end{subequations}
where $\structcost = \mathbb{E}_{\world, \sigma}[\cost^{1}(\var^1(\sigma), \var^2(\sigma,\world))]$ for brevity, and \cref{structure_limits} can be interpreted as encoding \ac{up}'s limited uncertainty-reducing resources, e.g., a limited number of security cameras.

\subsection{Two-Stage Problem}

Our goal in this work is to develop a decision-making algorithm for a \ac{up} aiming to optimally allocate information-gathering assets,
and then optimally play a two-player non-cooperative game where the other player's costs are unknown.
We now have all the parts to model this scenario as a two-stage problem composed of a signal structure selection problem \cref{structure_selection} and the smooth game described in \cref{sec:smooth}.
Given a prior distribution over the worlds $p(\world)$, \ac{up} solves two stages: 
\begin{itemize}
    \item Stage 1: Solve \cref{structure_selection} to obtain a signal structure $\structs$.
    \item Stage 2: Given a signal structure, assemble a policy that maps signals to \ac{up} decisions, i.e., $\var^1: \signals \to \feasibleset^1$ using the solution to \cref{game_signal}.
\end{itemize}
For both stages, we seek locally optimal solutions.
\section{Solving the Two-Stage Problem}

\begin{algorithm}[bp]
\SetAlgoLined
\DontPrintSemicolon
\caption{Solve Two-Stage Problem}
\label{alg:solve_two_stage}
{
\small
    \textbf{Input:} $p(\world)$, step size $\alpha \in \mathbb{R}$, initial guess $\structs_0$.\\
    $\structs \gets \structs_0$\\
    \While{$!\mathrm{converged}$}{
        $\var^1, \var^2 \gets \mathrm{solve Stage 2}(\structs, p(\world))$\\
        $\frac{d\structcost}{d\structs} \gets \mathrm{composeDerivative}(\structs, \var^1, \var^2)$\\
    $\structs_{0} \gets \structs - \alpha \frac{d\structcost}{d\structs}$\\
    $\structs \gets \mathrm{project}(\structs_{0})$ \tcp{Onto constraints}
}
$\var^1, \var^2 \gets \mathrm{solve Stage 2}(\structs, p(\world))$\\
    \Return{$\structs,\var^1, \var^2$}
}
\end{algorithm}

\Cref{alg:solve_two_stage} summarizes our approach for solving this two-stage problem. At a high level:
(i) we solve the two-stage problem by making an initial guess for $\structs$, 
(ii) then, we use this guess to solve Stage 2 in $\cref{game_signal}$, 
and 
(iii) finally, we descend the gradient $\frac{d\structcost}{d\structs}$ and project the resulting value of $\structs$ onto the simplex constraints \cref{structure_probability} and \cref{structure_limits}.

As \cref{alg:solve_two_stage} depends upon the gradient $\frac{d\structcost}{d\structs}$, we first discuss the existence and computation of this derivative.
The Stage 1 objective $\structcost$ from \cref{structure_objective} is a function of the decision variables for Stage 2, i.e., $\var^1(\signal), \var^2(\signal,\world)$. 
The value of these variables depends on the Stage 1 signal structure selection via \cref{eq:game_signal_a}.
Therefore, when computing $\frac{d\structcost}{d\structs}$ we must consider the relationship between the solution of the Stage 2 problem \cref{game_signal} and changes in $\structs$. 
Proceeding formally, we compute the total derivative of $\structcost$ with respect to $\structs$ as
\begin{align}
\label{eq:dKdr}
    \frac{d\structcost}{d\structs}
    = \nabla_{\structs} \structcost &+ \nabla_{\vars} \structcost \nabla_{\structs} \vars\\
    \nonumber = \nabla_{\structs} \structcost
    &+ \sum_{j=0}^\dimworlds \nabla_{\var^1(j)} \structcost \, \nabla_{\structs} \, \var^1(j)
     \\  &+ \sum_{j,i=0}^\dimworlds \nabla_{\var^2(j,\world_i)} \structcost \, \nabla_{\structs} \, \var^2(j,\world_i)
\end{align}
Thus, computing $\frac{d\structcost}{d\structs}$ requires us to compute $\nabla_{\structs} \, \vars$, which is the derivative of a 
Nash equilibrium solution with respect to parameters of players' objectives, which may not be well-defined, in general.
Thus, we offer the following proposition that provides sufficient conditions for the existence of $\frac{d\structcost}{d\structs}$ in unconstrained games with $\feasibleset^i = \mathbb{R}^n, \forall i \in \{1,2\}$. 
In the experimental section we describe how our solver is extended to the constrained setting. 

\begin{proposition}
\label{prop:derivative}
    Let $\structs$ be a point in the relative interior of the simplex, and let $(\vars^{1*},\vars^{1*})$ be a Nash equilibrium solution for an associated Stage 2 game with no constraints ($\mathcal{X} = \mathbb{R}^n$).
    Then, the gradient $\frac{d\structcost}{d\structs}$ exists at $\structs$ if the following conditions hold:
    \begin{enumerate}
        \item \label{prop:derivative.assumpt_1} $\mathbb{E}_{\world|0}[\cost^{1}(\varbrev^{1}, \varbrev^{2};\world)]$ and $\cost^{2}(\varbrev^{1}, \varbrev^{2}_1;\world_k)$, $k \in [m]$ are twice-differentiable with respect to $\varbrev^1$ and $\varbrev^2_k$, respectively.
        \item \label{prop:derivative.assumpt_2} $\nabla^2_{\varbrev^1}\mathbb{E}_{\world|0}[\cost^{1}(\varbrev^{1}, \varbrev^{2};\world)]$ and each $\nabla^2_{{\varbrev^2_k}}\cost^{2}(\varbrev^{1}, \varbrev^{2}_1;\world_k)$, $k=[\dimworlds]$, are invertible.
        \item \label{prop:derivative.assumpt_3} The matrix $\mathbf{E}$ given by 
        $$\mathbf{E} = \\\nabla^2_{\varbrev^1} \mathbb{E}[\cost^1] 
        - \sum_{i=1}^\dimworlds \nabla_{\varbrev^2_i, \varbrev^1} \mathbb{E}[\cost^1](\nabla^2_{\varbrev^{2}_i}\cost^{2})^{-1}\nabla_{\varbrev^1,\varbrev^2_i} \cost^2$$
        is invertible.
    \end{enumerate}
    where we employ the notation $\varbrev^1 = \var^{1*}(0)$ and $\varbrev^2_i = \var^{2*}(0, \world_i)$ for brevity.
    All the matrices are evaluated at $(\structs,\vars^{1*},\vars^{1\star})$
\end{proposition}

\begin{proof}

    Stage 2 can be decoupled into a set of perfect information games with signals $\signal>0$---i.e., \cref{eq:perfect_info_p1_start,eq:perfect_info_p1_end,eq:perfect_info_p2_start,eq:perfect_info_p2_end}---and a single imperfect information game, where $\signal = 0$---i.e., \cref{eq:game_signal_a,eq:imperfect_info_start,eq:imperfect_info_end}. 
    Due to \cref{assumption:no_false_positives}, only the imperfect information game depends on the signal structure selection $\structs$, and only these solutions need to be considered when computing $\nabla_\structs \structcost$.
    That is, $\nabla_{\structs} \, \var^1(j)=0$ and $\nabla_{\structs} \, \var^2(j,\world_j)=0$ for all $j>0$, and thus Eq.~\ref{eq:dKdr} reduces to
    \begin{align}
        \label{eq:dKdr.proof}
        \frac{d\structcost}{d\structs} &= \nabla_\structs\structcost + \underbrace{\nabla_{\varbrev^1}\structcost\nabla_\structs\varbrev^1 + \sum_{j=1}^\dimworlds\nabla_{\varbrev^2_j}\structcost \nabla_{\structs}\varbrev^2_j}_{\nabla_{\solution}\structcost \nabla_\structs\solution}
    \end{align}
    where $\solution \triangleq (\varbrev^1, \varbrev^2_1, \hdots, \varbrev^2_\dimworlds)$.
    The terms $\nabla_\structs \structcost$ and $\nabla_\solution \structcost$ are straight-forward to compute due to Assumption~\ref{prop:derivative.assumpt_1}. 
    However, $\nabla_\structs \solution$ requires more work since it is not immediately clear how to differentiate the solution of the Stage 2 subgame \cref{game_signal}. 

    Fortunately, we can find an expression for $\nabla_\structs \solution$ by analyzing the structure of the game solutions and applying the implicit function theorem \cite{dontchev2009implicit}. 
    First, we define a solution to the Stage 2 game as a Nash equilibrium. 
    Then, using the fact that a Nash equilibrium implies first-order stationarity for all players, we leverage the implicit function theorem and derive sufficient conditions for the existence of $\nabla_\structs \solution$.

    Solutions to the Stage 2 game \cref{game_signal} are points that satisfy Nash equilibrium conditions where no player has a unilateral incentive to deviate from their chosen strategy. 
    Concretely, in a game with $\numplayers$ players, strategy $\vars^* = [\var^{1*}, \dots, \var^{\numplayers*}]$ is a Nash equilibrium if it satisfies 
    \begin{equation}
        \label{nash}
        \cost^i(\vars^*) \leq \cost^i([\var^{i}, \var^{-i*}])~\forall i \in [\numplayers],
    \end{equation}
    where $[\var^{i}, \var^{-i*}]$ denotes a strategy where only player $i$ deviates from $\vars^*$.
 
    Inequality \cref{nash} implies that first-order stationarity must hold for all players, i.e.,  
    \begin{equation}
        \gradkkt = 
        \begin{bmatrix}
            \nabla_{\varbrev^1}\mathbb{E}_{\world|0}[\cost^{1}(\varbrev^{1*}, \varbrev^{2*};\world)]  \\ 
            \nabla_{\varbrev^{2}_1}\cost^{2}(\varbrev^{1*}, \varbrev^{2*}_1;\world_1)\\
            \vdots\\
            \nabla_{\varbrev^{2}_\dimworlds}\cost^{2}(\varbrev^{1*}, \varbrev^{2*}_\dimworlds;\world_\dimworlds)
        \end{bmatrix} 
        = \mathbf{0}.
        \label{eq:gradkkt}
    \end{equation}
    Then, assuming certain regularity conditions are met, by the implicit function theorem \cite{dontchev2009implicit}[Thm. 1B.1],
    there exists a localized solution map $\solution: \structs' \mapsto \solution(\structs')$ for any $\structs' \in U$, a neighborhood of $\structs$ on which $\gradkkt(\solution(\structs'),\structs') = 0$.
    Moreover, for this localized solution map,
    $\nabla_\structs \solution$ exists and is given by
    \begin{equation}
        \nabla_\structs \solution = \nabla_\solution \gradkkt^{-1}\nabla_\structs \gradkkt.
    \end{equation}
    The following are the sufficient conditions for the existence and differentiability of this solution map in $U$: 
    \begin{enumerate}
        \item $\gradkkt = \mathbf{0}$
        \item $\gradkkt$ is continuously differentiable
        \item $\nabla_\solution \gradkkt$ is invertible.
    \end{enumerate}
    The first condition is already given by \cref{eq:gradkkt}.
    The second condition follows if we assume costs are twice-differentiable with respect to their respective decision variables $\varbrev^1$ and $\varbrev^2_i$ (Condition 1 in \cref{prop:derivative}).
        
    To show that the third condition is met we compute $\nabla_\solution \gradkkt$ and analyze its structure.  
    We begin by noting that
    \begin{equation}
        \nabla_\solution \gradkkt = \left[\begin{smallmatrix}
            \nabla^2_{\varbrev^1} \mathbb{E}[\cost^1] & \nabla_{\varbrev^2_1, \varbrev^1} \mathbb{E}[\cost^1] & \cdots & \nabla_{\varbrev^2_\dimworlds, \varbrev^1} \mathbb{E}[\cost^1] \\
            \nabla_{\varbrev^1, \varbrev^2_1} \cost^2_1 &  \nabla^2_{\varbrev^2_1} \cost^2_1 & \cdots & \nabla_{\varbrev^1_0, \varbrev^2_\dimworlds} \cost^2_1\\
            & \vdots & & \\
            \nabla_{\varbrev^1, \varbrev^2_\dimworlds} \cost^2_\dimworlds &  \nabla_{\varbrev^2_1, \varbrev^2_\dimworlds} \cost^2_\dimworlds & \cdots & \nabla^2_{\varbrev^2_\dimworlds} \cost^2_\dimworlds
        \end{smallmatrix}
        \right].
    \end{equation}
    where $\nabla_{u,v} = \nabla_u \nabla_v$, $J^i_k = J^i( \cdot ; \omega_k)$ and we omit the other arguments for brevity.
    Decisions made by Player 2 are independent for all worlds. Therefore, $\nabla_{\varbrev^2_i, \varbrev^2_j} \cost^2_j = \mathbf{0}, i \neq j$
    and all off-diagonal $\dimvar \times \dimvar$ blocks in the bottom-right $\dimvar\dimworlds \times \dimvar\dimworlds$ block of $\nabla_\solution \gradkkt$ are zero.
    \begin{equation}
        \implies \nabla_\solution \gradkkt = 
        \left[\begin{smallmatrix}
            \nabla^2_{\varbrev^1} \mathbb{E}[\cost^1] & \nabla_{\varbrev^2_1, \varbrev^1} \mathbb{E}[\cost^1] & \hdots & \nabla_{\varbrev^2_\dimworlds, \varbrev^1} \mathbb{E}[\cost^1] \\
            \nabla_{\varbrev^1,\varbrev^2_1} \cost^2 &  \nabla^2_{\varbrev^2_1} \cost^2 & \hdots & \mathbf{0}\\
            \vdots & \vdots & \ddots & \vdots \\
            \nabla_{\varbrev^1, \varbrev^2_\dimworlds} \cost^2 &  \mathbf{0} & \hdots & \nabla^2_{\varbrev^2_\dimworlds} \cost^2
        \end{smallmatrix}
        \right]
    \end{equation} 
    We can re-write $\nabla_\solution \gradkkt$ as the block matrix
    \begin{equation}
        \nabla_\solution \gradkkt =
        \begin{bmatrix}
            \mathbf{A} & \mathbf{B} \\
            \mathbf{C} & \mathbf{D}
        \end{bmatrix}
    \end{equation}
    where
    \begin{subequations}
        \begin{align}
            \mathbf{A} &= \nabla^2_{\varbrev^1} \mathbb{E}[\cost^1]\\
            \mathbf{B} &= \left[\nabla_{\varbrev^2_1, \varbrev^1} \mathbb{E}[\cost^1]\,\hdots\, \nabla_{\varbrev^2_\dimworlds, \varbrev^1} \mathbb{E}[\cost^1]\right]\\
            \mathbf{C} &= 
            \begin{bmatrix}
                \nabla_{\varbrev^1,\varbrev^2_1} \cost^2_1\\
                \vdots\\
                \nabla_{\varbrev^1, \varbrev^2_\dimworlds} \cost^2_\dimworlds 
            \end{bmatrix}\\
            \mathbf{D} &= 
            \left[\begin{smallmatrix}
                \nabla^2_{\varbrev^2_1} \cost^2_1 & \hdots & \mathbf{0}\\
                \vdots & \ddots & \vdots \\
                \mathbf{0} & \hdots & \nabla^2_{\varbrev^2_\dimworlds} \cost^2_\dimworlds
            \end{smallmatrix}
            \right].
        \end{align}
    \end{subequations}
    If Player 1 and Player 2's cost Hessians are invertible (Condition 2), so are $\mathbf{A}$ and $\mathbf{D}$, and we can express $\det(\nabla_\solution \gradkkt)$ in terms of the Schur complement $\nabla_\solution \gradkkt / \mathbf{D}$: 
    \begin{equation}
        \det(\nabla_\solution \gradkkt) = \det(\mathbf{A})\det(\nabla_\solution \gradkkt / \mathbf{D}), 
    \end{equation}
    where $\nabla_\solution \gradkkt / \mathbf{D} = \mathbf{A} - \mathbf{B}\mathbf{D}^{-1}\mathbf{C}$. 
    Then, $\nabla_\solution \gradkkt$ is invertible if and only if $\det(\nabla_\solution \gradkkt / \mathbf{D}) \neq 0$ (since $\det(\mathbf{A}) \neq 0$).  

    If we let $\mathbf{E} = \nabla_\solution \gradkkt / \mathbf{D}$ we have
    \begin{equation}
        \mathbf{E} = \nabla^2_{\varbrev^1} \mathbb{E}[\cost^1] - \sum_{j=1}^\dimworlds \nabla_{\varbrev^2_j, \varbrev^1} \mathbb{E}[\cost^1](\nabla^2_{\varbrev^{2}_j}\cost^{2}_j)^{-1}\nabla_{\varbrev^1,\varbrev^2_j} \cost^2_j.
    \end{equation}
    Thus, $\nabla_\solution \gradkkt$ is invertible if and only if $\mathbf{E}$ is invertible (Condition 3).

    Therefore, by the implicit function theorem, $\nabla_\structs \solution$ exists if Conditions 1-3 hold.
    Moreover, using \cref{eq:dKdr.proof}, it follows that $\frac{d \structcost}{d \structs}$ exists as well.
\end{proof}

\section{Experimental Results}
\label{sec:experiments}
In this section we demonstrate our proposed formulation with a zero-sum tower-defense game. 
We include a variety of experiments meant to give intuition about the Stage 1 cost landscape and the optimal selection given a selected signal structure. 
We also present a visualization of the output of \cref{alg:solve_two_stage}.
Code for all implementations will be made publicly available. 

We define a zero-sum tower-defense game in which the tower, referred to as the defender, may be attacked from one of three directions by an attacker.
In Stage 1, the defender (\ac{up}) allocates scouts in each direction, corresponding to the selection of a signal structure.
These scouts are meant to warn the defender of the preferred attack direction.
In Stage 2, the defender receives a signal from its scouts, represented as an integer $\sigma \in \{0, 1,2, 3\}$. 
Integers $1 \leq \sigma \leq 3$ inform the defender what the world is, and in every world $\world \in \worlds$, the attacker has a preferred attack direction.
A zero signal corresponds to the case where the defender gets no warning about the world value. 
After receiving a signal, the defender plays a zero-sum game with the defender using a policy $\var^1: \signals \to \feasibleset^1$ that maps signals to defense allocations.
Similarly, the attacker plays the game with a policy $\var^2: \signals \times \worlds \to \feasibleset^2$, that maps the defender signal (which we assume the attacker knows) and the world (known to the attacker), to attack allocations. 

Stage 1 consists of the signal structure selection problem defined in \cref{structure_selection} with defender cost function $\cost^1$ given by
\begin{subequations}
    \label{defender_cost}
    \begin{align}
        \cost^1(\var^1(i), \var^2(i,\world); \world) = -\cost^2(\var^1(i), \var^2(i,\world); \world)
    \end{align}
\end{subequations}
where $\cost^2$ will be defined shortly.
Each world $\world$ corresponds to an attacker cost function with a preference towards a particular direction of attack.

In Stage 2, the policies $\var^1, \var^2$ are assembled from the solution of the Stage 1 game \cref{game_signal} with $\cost^1$ and $\cost^2$ as defined in this section.
They are also constrained such that $\feasibleset^i = \{\var^i ~|~ \sum_{j=0}^\dimvar\var^i_j = 1, \var^i \geq \mathbf{0} \}$.

The attacker cost function is given by
\begin{subequations}
    \label{attacker_cost}
    \begin{alignat}{2}
        & \cost^2(\var^1(i),\, &&\var^2(i,\world),\world) = -\sum_{j=1}^\dimworlds\activation(\delta_j(\world)) \; \delta^2_j(\world)\\
        \label{attack_defense_delta}
        &\text{where} &&\difference_j(\world) = \preference_j(\world) \var_j^2(i)- \var_j^1(i,\world)\\
        \label{eq:activation}
        & &&\activation(\difference) = \frac{1}{1 + e^{-2k\difference}}\\
        \label{preference_condition}
        & &&0 < \preference_k(\world_j) < \preference_j(\world_j), \forall k\neq j.
    \end{alignat}
\end{subequations}
The term $\difference_j(\world)$ in \cref{attack_defense_delta} is the difference between a scaled attacker's allocation $\var_j^2(i)$ and the defender allocation $\var_j^1(i)$ in direction $j$.
Informally, a large $\difference_j(\world)$ corresponds to a mismatch between defense and attack allocations in direction $j$, a win for the attacker.
The scaling effect of $\preference_j(\world)$ can be interpreted as a force multiplier in each direction of attack. 
Inequality \cref{preference_condition} implies that $\preference_k(\world_j)$ is larger when $k=j$. 
Therefore, it is easier for the attacker to gain a numerical advantage in that direction.
This is how we encode a direction preference, since the cost in (\ref{attacker_cost}a) is decreasing in each $\delta_j$.
The cost function in \cref{eq:activation} uses a logistic factor with sharpness parameter $k$ that ``activates'' only if $\difference > 0$. 
This ensures that the attacker is not penalized for situations where the defender allocates resources in a direction where the attacker is not present. 

We gather the preferences for each world and direction in the matrix $\preferences$ as follows:
\begin{equation}
    \label{preference_matrix}
    \preferences = 
    \begin{bmatrix}
        \preference^\top(\world_1)\\
        \preference^\top(\world_2)\\
        \preference^\top(\world_3)
    \end{bmatrix}
    = 
    \begin{bmatrix}
        \preference_1(\world_1) & \preference_2(\world_1) & \preference_3(\world_1)\\
        \preference_1(\world_2) & \preference_2(\world_2) & \preference_3(\world_2)\\
        \preference_1(\world_3) & \preference_2(\world_3) & \preference_3(\world_3)\\
    \end{bmatrix}.
\end{equation}
In this case, \cref{preference_condition} implies that the diagonal elements are the largest for every row. 

This experiment can be understood as a version of the Colonel Blotto game with smooth payoff functions.
In the standard formulation of a Colonel Blotto game, two players must simultaneously allocate forces across a set of battlefields \cite{roberson2006blotto}. 
At every battlefield, the player with the largest number of forces wins, and the payoff for all players depends on the number of battlefields they win.
By contrast, ours is a game of degree, as there is a continuously varying quantity that may be won from each battlefield.



Computing the derivative for this experiment requires considering the effects of the constraints on the solution of Stage 2 as these constraints were not considered in \cref{prop:derivative}.
As shown in \citet{liu2023learning}, when strict complementarity holds, and under mild assumptions, the derivative $\nabla_\structs \solution$ can be uniquely computed. 
For the case of weak complementarity, the derivative is not well-defined, but subgradients can still be computed.
Our implementation of \cref{alg:solve_two_stage} accounts for both cases.

\subsection{Understanding the Cost Landscape}
We begin by plotting the Stage 1 cost as a function of scout allocation. 
To do so, we first select a valid scout allocation $\structs$ and solve \cref{game_signal} to obtain $\var^1(\signal)$ and $\var^2(\signal, \world), \forall \world \in \worlds, \signal \in \signals$ for which $p(\signal, \omega)>0$.
Then, we use these variables to compute the Stage 1 cost \cref{structure_objective}.  
Note that $\structs$ belongs to the 2-simplex, and therefore, has only two degrees of freedom. 
This means we can plot the Stage 1 cost \cref{structure_objective} as a 3-D surface where the $(x,y)$-axes are $\structs_1$ and $\structs_2$, and the $z$-axis is the normalized expected cost.
We normalize the costs by dividing with the highest observed cost among all $\structs$ in the simplex. 

For this experiment, we start with a uniform prior $p(\world_i) = \frac{1}{3}, \, \forall i \in \{1,2,3\}$, a sharpness parameter $k = 10.0$, and the following attacker preference matrix: 
\begin{equation}
    \label{preference_matrix_example}
    \preferences = 
    \begin{bmatrix}
        3.0 & 2.0 & 2.0\\
        2.0 & 3.0 & 2.0\\
        2.0 & 2.0 & 3.0\\
    \end{bmatrix}.
\end{equation}

\begin{figure}[h!]
    \centering
    \includegraphics[width=0.80\linewidth]{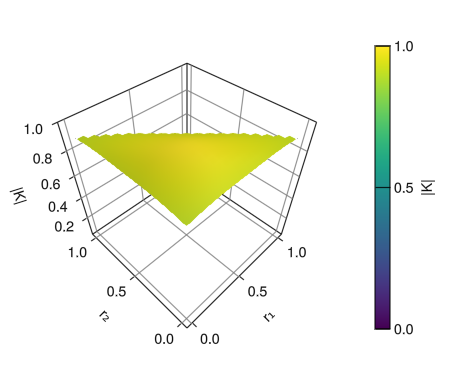}
    \caption{Normalized Stage 1 cost $|K|$ as a function of scout allocation $\structs$.}
    \label{fig:stage_1_cost}
\end{figure}

The resulting plot in \cref{fig:stage_1_cost} reveals the surprising fact that, for this parameter regime, the Stage 1 cost landscape is relatively flat. 
This implies that scout allocation has only a small effect on the expected cost for the defender at Stage 1. 

\begin{figure}[tp]
    \centering
    \includegraphics[width=0.85\linewidth]{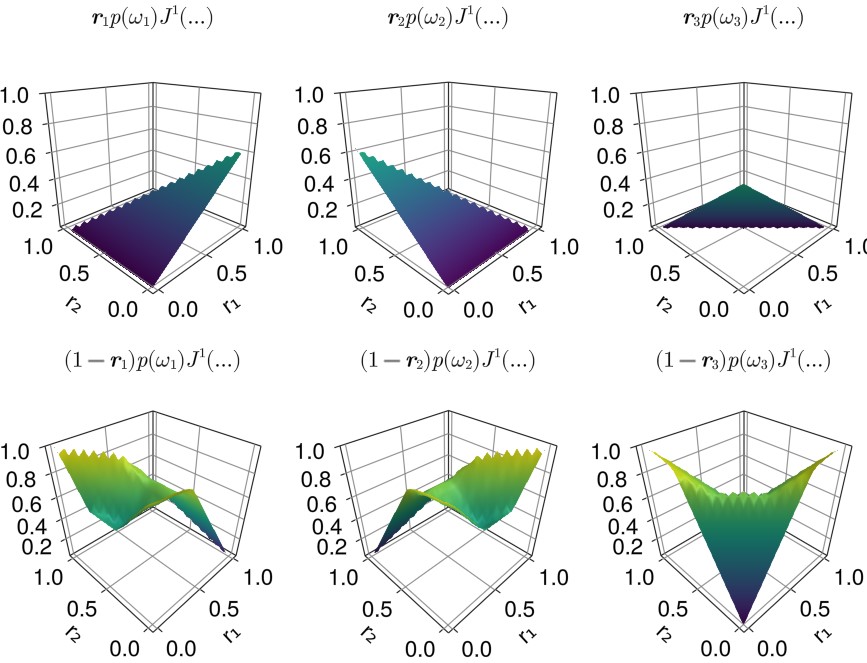}
    \caption{Values for each term in \cref{stage_1_perfect_info_terms} as a function of $\structs$. Normalized by the highest value observed across all six terms.}
    \label{fig:stage_1_terms}
\end{figure}

To understand why, we begin by separating the Stage 1 cost into the six terms from which it is composed:
\begin{subequations}
    \begin{align}
        \mathbb{E}_{\world,\signal}[&\cost^{1}(\var^1(\signal), \var^2(\signal,\world); \world)] =\\
        &\sum_{i=1}^3\structs_i \prior(\world_i)\cost^{1}(\var^1(i), \var^2(i,\world_i)) + \\ \label{stage_1_perfect_info_terms}
        &\sum_{i=1}^3(1 - \structs_i) \prior(\world_i)\cost^{1}(\var^1(0), \var^2(0,\world_i)).
    \end{align}
\end{subequations}
Then, we plot the separate contribution of each term in \cref{fig:stage_1_terms}, where we readily see how the structure of the individual terms results in a cancellation effect that leads to the relatively flat Stage 1 cost in \cref{fig:stage_1_cost}.

The result is that the optimal scout allocations are found at the vertices of the simplex. 
This result has an interesting interpretation: it is in the defender's best interest to completely remove uncertainty about one world, instead of distributing its information-gathering resources across many directions. 

\subsection{Corresponding Attack and Defense Decisions}

\begin{figure}[h!]
    \centering
    \includegraphics[width=0.85\linewidth]{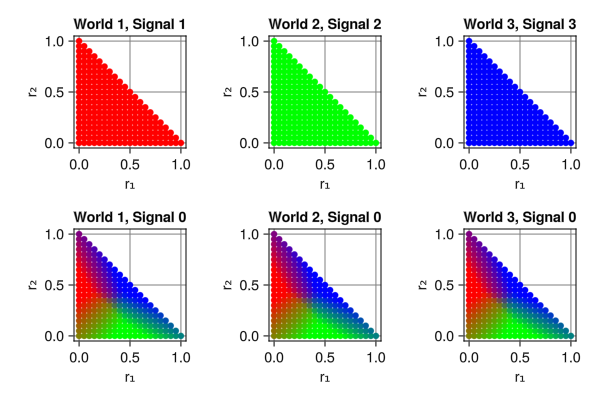}
    \caption{Defender policy $\var^1$ as a function of $\structs$ and signal/world.}
    \label{fig:stage_1_controls_defense}
\end{figure}

We now present how players' decisions change as a function of signal structure selection $\structs$.
To this end, in \cref{fig:stage_1_controls_defense} we color each point in the $\structs$ simplex with an RGB color whose component intensity is given by a Stage 2 decision variable. 
For example, given an $\structs$, if the defender's optimal Stage 2 decision for $\signal = 0$ is $\var^1(0) = [0.5, 0.5, 0.0]^\top$, then point $\structs$ is colored by the (normalized) RGB triplet $(0.5, 0.5, 0.0)$.
This figure illustrates what the defense policy $\var^1$ is for every possible combination of signal and signal structure selection. 

\begin{figure}[h!]
    \centering
    \includegraphics[width=0.85\linewidth]{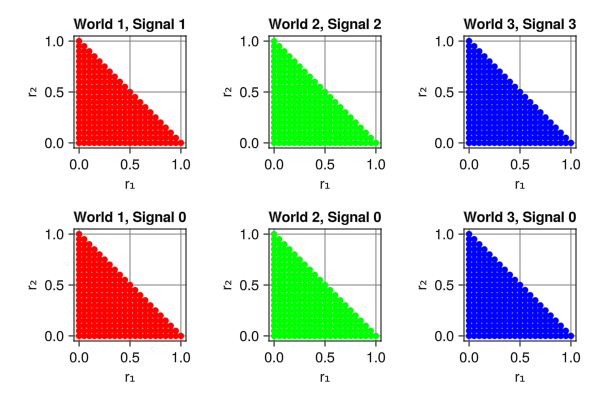}
    \caption{Attacker policy $\var^2$ as a function of $\structs$ and signal/world.}
    \label{fig:stage_1_controls_attack}
\end{figure}

 We note that in the case of detection ($\signal \neq 0$, top three plots in \cref{fig:stage_1_controls_defense}), the defense policy is to allocate all of the resources in the direction of the highest attack preference. 
 For example, consider the top-middle plot in \cref{fig:stage_1_controls_defense} which displays the defender's decisions when $(\world, \signal) = (\world_2, 2)$. 
 The preference matrix \cref{preference_matrix} implies that in $\world_2$, the attacker prefers direction 2. 
 The resulting optimal defensive action is to allocate all resources in that direction, as shown by the green surface with RGB triplet $(0.0, 1.0, 0.0)$. 
 To help understand why, we examine the attack policy, shown in  \cref{fig:stage_1_controls_attack}, for the same world and signal combination.
 Note that the surface in the top-middle plot is all green, which means that the attacker allocates all of its resources in direction 2.
 The interpretation is that even though a signal $\signal = 2$ informs the defender what the world is, and therefore, what the preferred attack direction is, the optimal choice for the attacker is still to attack in its preferred direction. 
 In response, the defender allocates all of its resources in that same direction. 
 This pattern holds for any of the complete information cases, i.e.,  $(\world, \signal) = (\world_i, i)$.
 This situation arises because the benefits conferred by a larger attack multiplier in the preferred direction are large enough to offset the actions taken by a defender who is aware of this preference. 

When the defender has no information about what the world is ($\signal = 0$), its decision will be based only on its knowledge of how it allocated information-gathering resources in Stage 1.
This may result in a mixed allocation in different directions. 
For example, if the defender's Stage 1 decision is $\structs = [1,0,0]^\top$ and it received a signal of $0$, then $p(0|\world_1) = 0$, implying that the value of the world must be $\world_2$ or $\world_3$.
Therefore, given a uniform prior and no other information, the optimal decision must be to distribute its resources between the preferred directions in $\world_2$ and $\world_3$.
This is seen in \cref{fig:stage_1_controls_defense} for $\signal=0$, with the selected decision for the bottom right corner ($\structs=[1,0,0]^\top$) represented in a teal color, which is a mixed allocation between directions 2 and 3 (green and blue, respectively).

\subsection{Making One World Riskier}

\begin{figure*}[h!]
    \centering
    \includegraphics[scale=0.3]{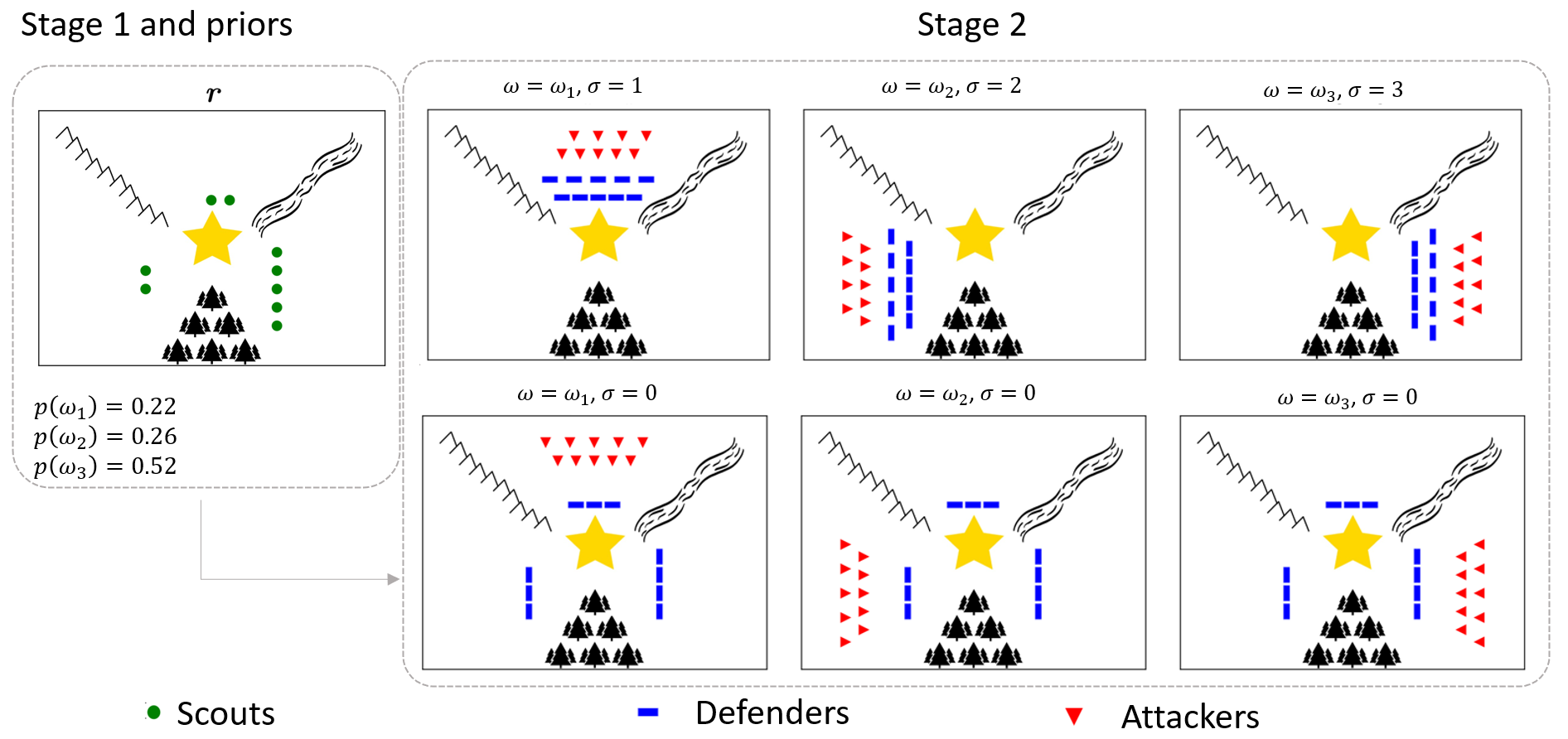}
    \caption{Visualization of a tower defense game for a given set of priors: $p(\world_{i}), \forall \world_i \in \worlds$. The scout allocations are depicted by green circles in Stage 1. The attacker and defender allocations are depicted by red triangles and blue rectangles, respectively. 
    In Stage 2, we depict the attacker and defender policies for all combinations of signals and worlds.}
    \label{fig:TD_visual}
\end{figure*}

We now examine how the decision landscape changes as we perturb the cost functions.
Let us consider the scenario in which one world is much riskier than the others---that is, for one of the worlds, the preferred direction has a larger force multiplier than the force multipliers of the preferred directions of the other worlds. 
Mathematically, we can model this by defining a perturbed preference matrix $\preferences_\perturbation$ as follows: 
\begin{equation}
    \label{eq:perturbed_preference}
    \preferences_\perturbation = 
    \begin{bmatrix}
        3.0 + \perturbation & 2.0 & 2.0\\
        2.0 & 3.0 & 2.0\\
        2.0 & 2.0 & 3.0\\
    \end{bmatrix}.
\end{equation}

\begin{figure}[h!]
    \centering
    \includegraphics[width=0.85\linewidth]{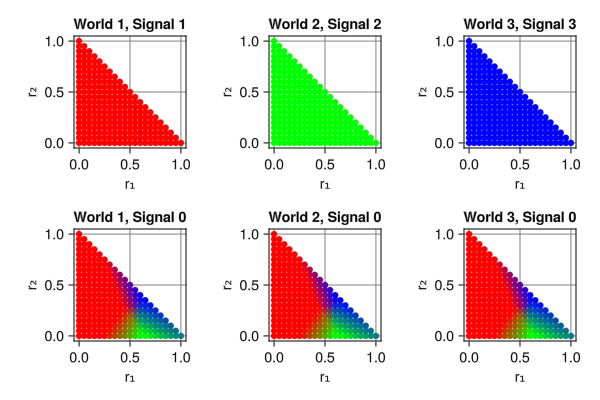}
    \caption{Defender policy $\vars^1$ given a perturbed preference matrix $\preferences_\perturbation$ \cref{eq:perturbed_preference} with $\perturbation = 2$. Red ``spreads out'' to the right side of the simplex, because the lower probabilities of the risky world (as one moves toward the right vertex of the simplex) are outweighed by the additional expected cost.}
    \label{fig:perturbed_controls}
\end{figure}

In \cref{fig:perturbed_controls} we present the resulting defender policy for the case where $\perturbation = 2$.
Compared to the original defender policy in \cref{fig:stage_1_controls_defense}, the defender policy remains the same for the complete information cases in the top three plots, i.e., when the world is detected and $(\world, \signal) = (\world_i, i)$. 
This makes sense because, within each world, the preferred attack direction has not changed, so once the world has been detected the best defense action remains unchanged whether $\perturbation = 0$ or not. 

On the other hand, in the incomplete information case ($\signal = 0$), the defense policy becomes more biased towards allocating defense resources in the risky direction. 
The expected cost associated with the risky world remains high enough to bias the minimizer of \cref{eq:game_signal_a} towards allocating resources in that direction---even if there is a low probability of that world given $\signal = 0$ (as given by the scout allocation $\structs$).

This explains the results shown in the bottom three plots of \cref{fig:perturbed_controls}, where a larger portion of the simplex is red when compared to \cref{fig:stage_1_controls_defense}.
Moreover, as one increases $\perturbation$, the red region expands over more of the $\structs$-simplex.
This is because the additional risk of $\world_1$ results in a situation where the defender's best decision is $\vars^1(0) = [1,0,0]^\top$, even when $p(\world_1 | 0)$ is low (which happens for $\structs$ closer to the right side of the simplex).
We note that the attacker policy remains unchanged and identical to the one seen in \cref{fig:stage_1_controls_attack}.

\subsection{Visualizing the Tower Defense Game}

Finally, in Figure~\ref{fig:TD_visual} we present a visualization of the output of \cref{alg:solve_two_stage} to help the reader gain an intuitive understanding of the relationship between the Stage 1 (``scouting'') and Stage 2 (``defending'') decisions.
We set the priors as shown, and compute the optimal $\structs$ value. This value corresponds to the allocation of scouts towards the north, east, and west directions --- with north being ``up,'' east ``right,'' and west ``left.''
The map of each world-signal $(\world, \signal)$ pair has three directions which correspond to the three available directions for attacking and defending.
\section{Conclusion}

We provide a two-stage, game-theoretic model for two-player noncooperative games where an uncertain player can pre-emptively gather information.
In the first stage, an uncertain player \acrfull{up} chooses how to allocate their information-gathering resources.
Critically, this model allows for a smooth resource allocation, in contrast to existing literature on the topic. 
In the second stage, \ac{up} receives information---also available to the \acrfull{cp}---in the form of a signal,
which \ac{up} and \ac{cp} use to play a noncooperative game.
We develop an algorithm that solves this two-stage problem, returning \ac{up}'s optimal allocation of information-gathering resources and \ac{up}'s policy that maps signals to actions in the noncooperative game.

This work can be extended in several directions. 
Within Blotto games, one could characterize how the cost landscapes change with different priors, cost parameters, and/or cost functions. 
One specific cost function of interest is the {\small\texttt{arctan}} payoff function introduced by \citet{Ferdowsi2018generalized} as part of the Generalized Blotto Game.
Other extensions include: (1) relaxing the assumptions on the signal structure, e.g., a game without the ``no-false-positive" \cref{assumption:no_false_positives}, (2) applying this to a game with higher dimensional decision variables and/or more worlds, and (3) playing stage 2 many times (perhaps infinitely), allowing for the \ac{up} to iteratively update their prior and improve outcomes. (4) A deeper analysis of conditions in \cref{prop:derivative}, and how they intuitively relate to the player objectives.
\section{Acknowledgements}

We would like to thank Michael Dorothy at the DEVCOM Army Research Laboratory and Yue Yu at the University of Texas at Austin for their valuable insights during the research and writing process. 

This research was sponsored by the Army Research Laboratory and was accomplished under Cooperative Agreement Number W911NF-23-2-0011.

\bibliographystyle{ACM-Reference-Format} 
\bibliography{fpe-ref}


\end{document}